# Semantic Segmentation Using Deep Learning to Extract Total Extraocular Muscles and Optic Nerve from Orbital Computed Tomography Images

**Fubao Zhu, Ph.D.[1], Zhengyuan Gao, B.S.[1], Chen Zhao, M.S.[2], Zelin Zhu, B.S.[1], Yanyun Liu, B.S.[1], Shaojie Tang, Ph.D.[3], Chengzhi Jiang, M.D.[4], Xinhui Li, M.D.[5], Min Zhao*, M.D.[5], Weihua Zhou*, Ph.D.[2]**

[1]School of Computer and Communication Engineering, Zhengzhou University of Light Industry, Zhengzhou, Henan, China

[2]College of Computing, Michigan Technological University, Houghton, MI, USA

[3]School of Automation, Xi'an University of Posts and Telecommunications, Xi'an, Shaanxi, China

[4]Department of PET-CT Center, Hunan Cancer Hospital, Changsha, China

[5]Department of Nuclear Medicine, Xiangya Hospital, Central South University, Changsha, China

**\*Corresponding authors:**

Weihua Zhou, PhD

College of Computing, Michigan Technological University

1400 Townsend Dr, Houghton, MI 49931

E-Mail: whzhou@mtu.edu

Min Zhao, MD

Department of Nuclear Medicine, Xiangya Hospital, Central South University

87 Xiangya Road, Changsha, Hunan, P.R. China, 410008

E-mail: mzhao1981@csu.edu.cn

**Abstract**

**Objectives:** Precise segmentation of total extraocular muscles (EOM) and optic nerve (ON) is essential to assess anatomical development and progression of thyroid-associated ophthalmopathy (TAO). We aim to develop a semantic segmentation method based on deep learning to extract the total EOM and ON from orbital CT images in patients with suspected TAO.

**Materials and Methods:** A total of 7,879 images obtained from 97 subjects who underwent orbit CT scans due to suspected TAO were enrolled in this study. Eighty-eight patients were randomly selected into the training/validation dataset, and the rest were put into the test dataset. Contours of the total EOM and ON in all the patients were manually delineated by experienced radiologists as the ground truth. A three-dimensional (3D) end-to-end fully convolutional neural network called semantic V-net (SV-net) was developed for our segmentation task. Intersection over Union (IoU) was measured to evaluate the accuracy of the segmentation results, and Pearson correlation analysis was used to evaluate the volumes measured from our segmentation results against those from the ground truth.

**Results:** Our model in the test dataset achieved an overall IoU of 0.8207; the IoU was 0.7599 for the superior rectus muscle, 0.8183 for the lateral rectus muscle, 0.8481 for the medial rectus muscle, 0.8436 for the inferior rectus muscle and 0.8337 for the optic nerve. The volumes measured from our segmentation results agreed well with those from the ground truth (all R>0.98, P<0.0001).

**Conclusion:** The qualitative and quantitative evaluations demonstrate excellent performance of our method in automatically extracting the total EOM and ON and measuring their volumes in orbital CT images. There is a great promise for clinical application to assess these anatomical structures for the diagnosis and prognosis of TAO.

## 1. Introduction

Thyroid-associated orbitopathy (TAO) is an autoimmune disorder characterized by inflammation and expansion of extraocular muscles (EOM), orbital fat and the lacrimal gland. It results in proptosis, corneal exposure, EOM motility impairment and optic neuropathy. Early diagnosis is crucial to the successful treatment of these pathologies. Clinical activity score (CAS) is the most widely used metric to evaluate inflammatory activity; however, it relies on subjective symptoms and patient cooperation. In addition, congestive changes of orbital soft tissues that are commonly presented in patients with severe TAO make it difficult to calculate CAS. Hence, accurate and robust assessment of inflammation is essential for determining the optimized treatment for TAO patients. Computer tomography (CT) is commonly used to measure the congestive changes in the extraocular muscles (EOM) and optic nerve (ON)[1]. Measured volumetric changes play an important role in the study of biophysical etiology, progression, and recurrence of TAO[2,3].

Precise segmentation of these anatomical structures is essential; however, it often requires the intervention of radiologists[4,5]. To reduce the workload of manual annotation and increase the reproducibility, several automatic segmentation methods based on templates[6], prior knowledge[7] and hand-crafted features[8] have been proposed. Nevertheless, the robustness and accuracy of these methods limit their clinical use. Recently, due to the success of deep learning models in a wide range of image processing applications, a substantial number of researches have aimed at developing medical image segmentation approaches using deep learning. The most recent advancements include end-to-end fully convolutional networks (CNN), such as U-Net[9] and V-Net[10]. In this paper, we propose a novel multi-class 3D V-Net (SV-Net), which contains several light-weight convolutional blocks with a limited number of parameters, to automatically extract total EOM and ON from orbital CT.

## 2. Materials and Methods

### 2.A. Image Acquisition and Description

Our study includes 97 suspected subjects with 7,879 orbital CT images obtained from Xiangya Hospital between August 2018 and June 2019. All patients were scanned with a 16-slice CT scanner (Precedence 16, SPECT/CT, Philips Medical Systems, Netherlands) without contrast enhancement. Orbital CT scan was obtained using contiguous axial slices, with the patient's head positioned parallel to the Frankfurt plane. Subjects were asked to close their eyes during image acquisition. The scanning parameters were as follows: 140 kV, 100 mA and 1-mm slice thickness. The spacing between slices is 0.5 mm and the mean in-plane pixel size is around 0.3393 mm × 0.3393 mm (0.22 mm-0.31 mm: 40 patients; 0.31 mm-0.40 mm: 32 patients; 0.39 mm-0.49 mm: 25 patients). Due to the size variation of eyes, the number of slices (Z direction) for each subject is different (45 to 126, mean 81). The image size in the Y direction varies from 142 to 642 (mean 291), but the size in the X direction is fixed as 768 in all subjects. The study was approved by the Ethics of Xiangya Hospital. Subjecs provided written informed consent for scan and participation in anonymized analyses.

## 2.B. Image Preprocessing and Data Augmentation

Figure 1 illustrates the workflow of image preprocessing. During preprocessing, one well-trained radiologist confirmed the start and end slices from the coronal view to ensure that the intermediate slices fully cover the total EOM and ON (Figure 1a). Regions of interest (ROI) with a uniform size of 256 × 256 were automatically generated from these immediate slices (Figures 1b and 1c), and then annotated by well-trained operators. Using an open-source interactive software tool Labelme[11], the contours of five regions, including superior rectus muscle, lateral rectus muscle, medial rectus muscle, inferior rectus muscle and optic nerve, were manually drawn in each slice as the ground truth (Figure 1c). Then, a one-hot encoding technique was used to map the annotated semantic maps into multi-channel semantic labels[12]. Data augmentation, such as translation, rotation, and Gaussian noise addition, was applied to the annotated images to increase the sample size of the training dataset, mitigate the overfitting problem, and improve the model robustness.

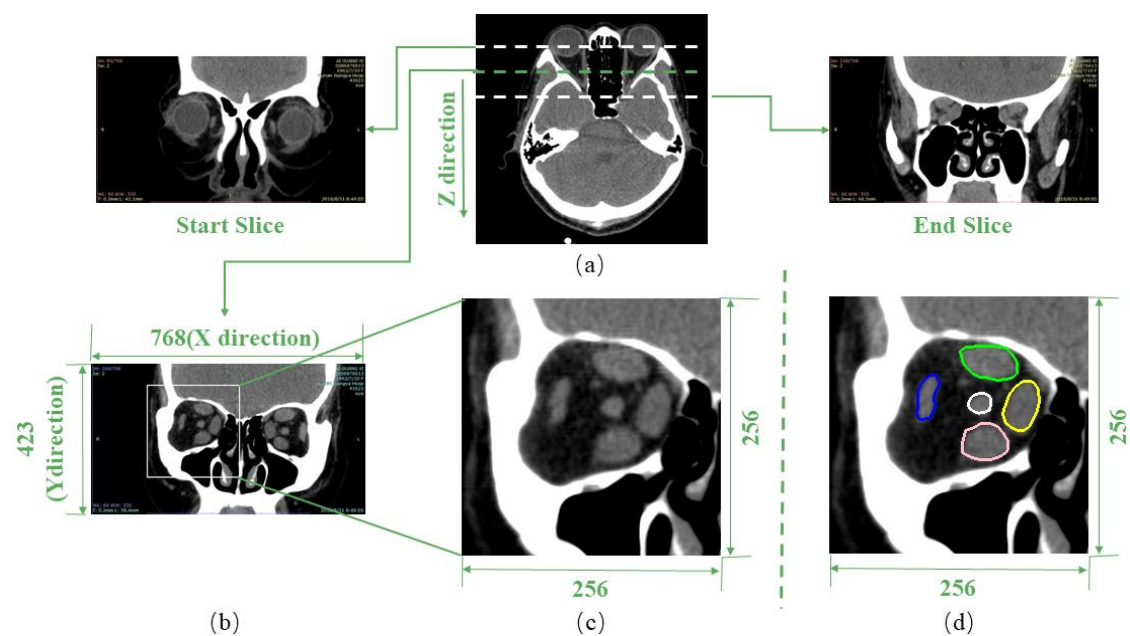

Figure 1. The workflow of image preprocessing. (a) Selecting the start and end slices. (b) A slice example. (c) A region of interest generated from (b). (d) The region of interest with annotations as the ground truth. Colors in the annotations: green, superior rectus muscle; blue, lateral rectus muscle; yellow, medial rectus muscle; pink, inferior rectus muscle; white, optic nerve.

## 2.C. Semantic Segmentation Network

Contrary to 2D semantic segmentation models, such as U-Net[9], we believe that the adjacent slices in our 3D orbital CT images provide supplementary information to assist segmentation with our prior knowledge on left ventricle[13] and proximal femur[14], so the proposed model for EOM and ON segmentation is a V-Net based deep neural network. The input of the model is a 3D cropped OCT image volume with a size of 256 × 256 × 32, and the output is a multi-channel semantic probability map. The architecture of the proposed semantic segmentation network is shown in Figure 2.

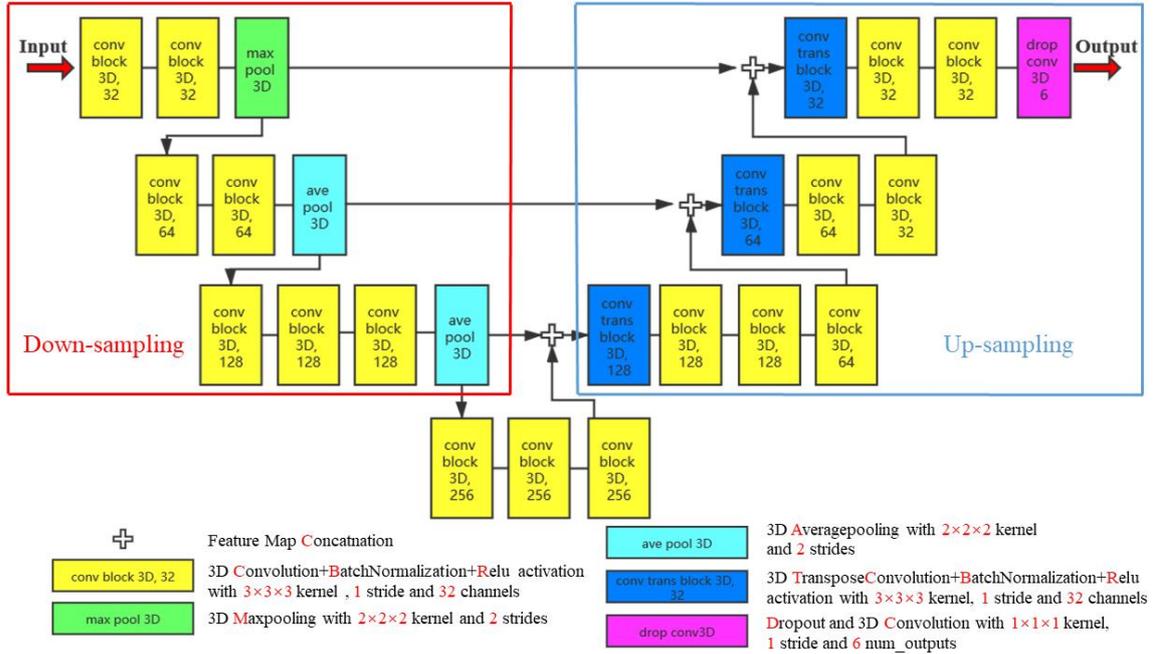

Figure 2. The architecture of our proposed semantic V-net network.

The proposed network consists of a down-sampling path extracting the features and an up-sampling path restoring the down-sampled features. The features are automatically extracted from the images through the convolutional operators during the down-sampling procedure.

Instead of using the convolutional layers with a kernel size of 5 × 5 × 5 in the V-net[10], the convolution layers containing a kernel size of 3 × 3 × 3 were adopted in our proposed model to reduce the number of weights in each layer, and the number of convolutional layers was reduced from 21 (in V-Net[10]) to 17. In addition, the number of weights in the network decreased from 9.6 M to 8.6 M, which significantly accelerated the training process.

In each convolutional layer, the pooling operators with a kernel size of 2 × 2 × 2 and a stride of 2 were applied to downsample the feature maps so that the receptive field in the deep network was large enough to extract high-level features. In addition, to avoid the gradient vanishing, accelerate network training, and guarantee the non-linearity of the network, batch normalization (BN)[15] and ReLU activation function were applied after creating each convolutional layer. At the end of the network, a convolutional layer with a kernel size of 1 × 1 × 1 was employed to convert the feature maps into multi-channel probability maps.

The objective function in this paper is a voxel-wise loss function based on a sigmoid cross-entropy. The definition is shown in Eqs. 1 and 2 for our SV-net,

$$p_{ijk} = \sigma\left(\text{logits}_{ijk}\right) = \frac{1}{1 + e^{-\text{logits}_{ijk}}} \tag{1}$$

$$\text{Loss} = \sum_i \sum_j \sum_k -[y_{ijk} \times \ln(p_{ijk}) + (1 - y_{ijk}) \times \ln(1 - p_{ijk})] \qquad (2)$$

where $\sigma$ is the sigmoid activation function, and $y_{ijk}$ represents the category of the ground truth in voxel (i,j,k). $\text{logits}_{ijk}$ indicates the predicted category of the voxel corresponding to the $y_{ijk}$.

## 2.D. Post-processing

To convert the multi-channel probability maps into binary maps in each channel, the category with the max probability of each voxel was selected as the predicted label. As illustrated in Figure 1, each target is a connected region without any holes. Thus, a hole filling algorithm[16] was applied to the binary maps in each channel. The number of voxels in each connected component was counted, and the largest connected region was selected as the final segmented result. Finally, we transformed each multi-channel binary map into six single-channel semantic maps corresponding to background, superior rectus muscle, lateral rectus muscle, medial rectus muscle, inferior rectus muscle and optic nerve, respectively.

## 2.E. Evaluation Methods

We randomly split the dataset into a training set consisting of 88 subjects with 7,143 images, and a testing set consisting of 9 subjects with 736 images. A 10-fold cross-validation was performed on the training set to select the optimal model. At the training stage, the 3D volume with 32 slices was randomly selected from the OCT sequence. To predict the entire 3D volume for each patient, we adopted a sliding window[17] scanning technique by cropping 32 slices from Z axis and feeding them into the network sequentially with a step of 1.

To test the effectiveness of the model, IoU, Specificity (SP) and Sensitivity (SN) were adopted. The definitions of IoU, SP and SN are shown in Eq. 3, Eq. 4 and Eq. 5, respectively. By taking the IoU of each class and averaging them, the mean IoU (mIoU) is calculated, as defined in Eq. 6.

$$\text{IoU}(R_k, R'_k) = \frac{|R_k \cap R'_k|}{|R_k| \cup |R'_k|} \qquad (3)$$

$$\text{SP}(R_k, R'_k) = \frac{|R_k \cap R'_k|}{|R'_k|} \qquad (4)$$

$$\text{SN}(R_k, R'_k) = \frac{|R_k \cap R'_k|}{|R_k|} \qquad (5)$$

$$\text{mIoU}(R, R') = \frac{1}{5} \sum_{k=1}^{5} \frac{|R_k \cap R'_k|}{|R_k| \cup |R'_k|} \qquad (6)$$

where the $R_k$ is the $k^{th}$ predicted image after the post-processing, and the $R'_k$ is the $k^{th}$ ground truth.

In addition, by considering the in-plane pixel size and slice thickness of OCT images, the volumes of the extracted total EOMs and ON were calculated. Then, root mean squared error (RMSE), mean absolute error (MAE), R value in Pearson analysis and relative error (RE) were employed to evaluate the volumes measured from our segmentation results against those from the ground truth.

## 3. Results

### 3.A. Parameter Setting

Our semantic segmentation model was implemented via TensorFlow 1.12.0 library with CUDA 9.0 and ran with a single Tesla P100 GPU with a 16GB GPU memory. To minimize the loss function defined in Eq. 2, an Adam optimizer[18] with a learning rate of 0.0001 was employed. In the proposed model, the batch size was set as 4 to fit the GPU memory. Each training batch took 1.68 seconds and the model converged within 125 epochs.

### 3.B. Experimental Results

As a comparison, a semantic U-Net[9] (SU-net), which replaced the 3D operations by 2D operations, was implemented as the benchmark model. We selected one slice from every 13 slices in the sequence of one subject as the representative examples to show the semantic segmentation results in the test dataset, as shown in Figure 3.

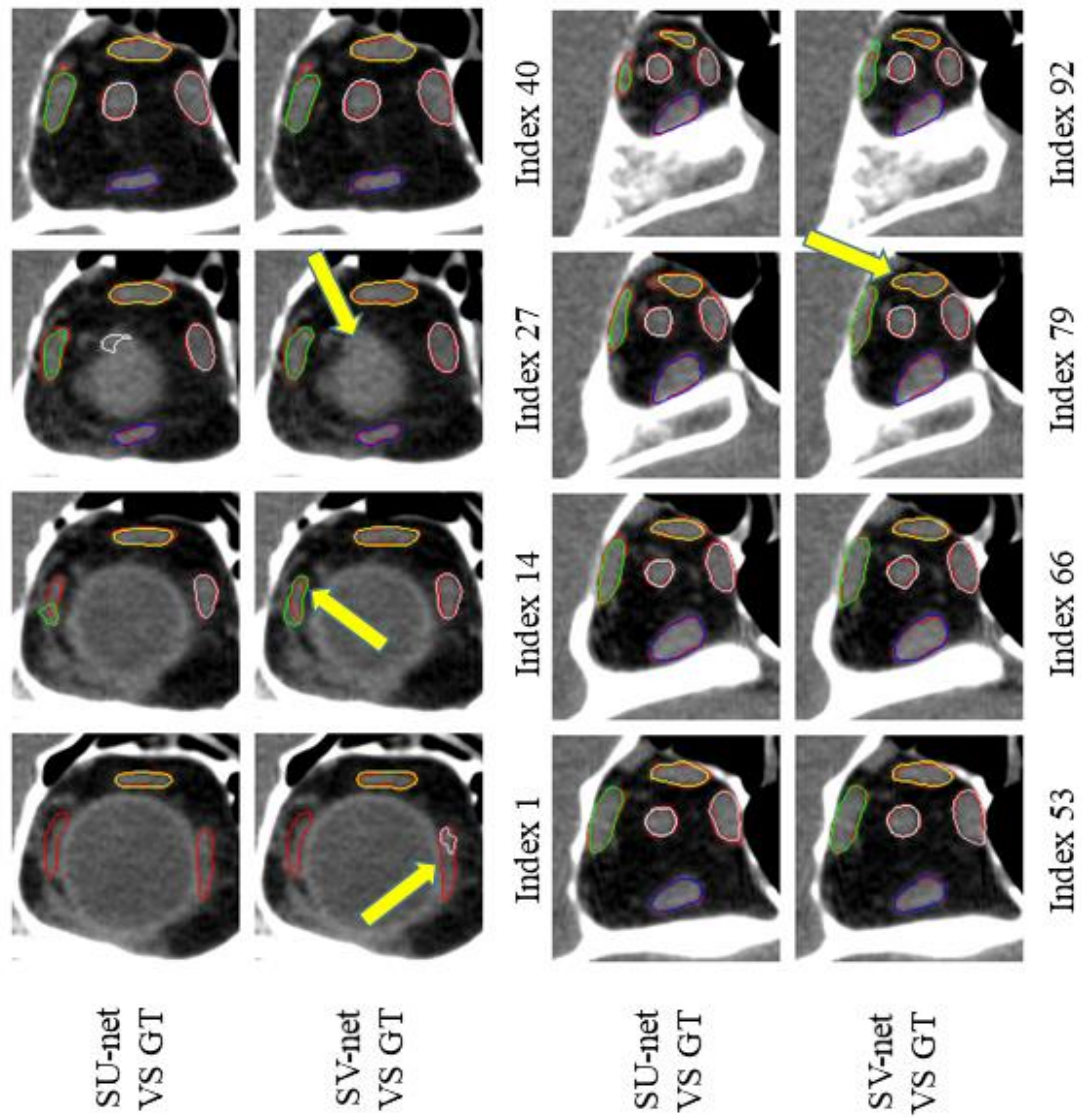

Figure 3. Semantic segmentation results. The red contours in all slices represent the ground truth. Colors in the annotations reflect the contours of segmentation results by the SU-net and SV-net: red, ground truth; green, superior rectus muscle; blue, lateral rectus muscle; yellow, medial rectus muscle; pink, inferior rectus muscle; white, optic nerve. SU-net, semantic U-net; SV-net, semantic V-net; GT, ground truth.

For the semantic information, both the SU-net and SV-net extracted the semantic information successfully in most slices. However, as highlighted by the yellow arrows, there are discrepancies between the segmentation results of SV-net and SU-net. Compared with SU-net, the segmentation results of SV-net are closer to the ground truth. It indicates that the 3D model achieved a higher performance with the spatial context from the 3D volume.

Table 1 shows the IoUs of the SU-net and SV-net. Our proposed SV-net achieved an overall mIoU of 0.8207, a SN of 0.9129 and a SP of 0.9996; an IoU of 0.7599, a SN of 0.8838 and a SP 0.9994 for superior rectus muscle; an IoU of 0.8183, a SN of 0.9105 and a SP 0.9996 for lateral rectus muscle; an IoU of 0.8481, a SN of 0.9276 and a SP 0.9996 for medial rectus muscle; an IoU of 0.8436, a SN of 0.9298 and a SP 0.9996 for inferior rectus muscle; an IoU of 0.8337, a SN of 0.9127 and a SP 0.9998 for ON.

Figure 4 shows the IoUs of our SV-net for the 5 regions in each test subject. Seven of all the 9 test subjects (77.8%) achieved an average IoU of greater than 0.80. Subject 8 obtained the maximum average IoU (0.8398), and Subject 7 produced the minimum average IoU (0.7904).

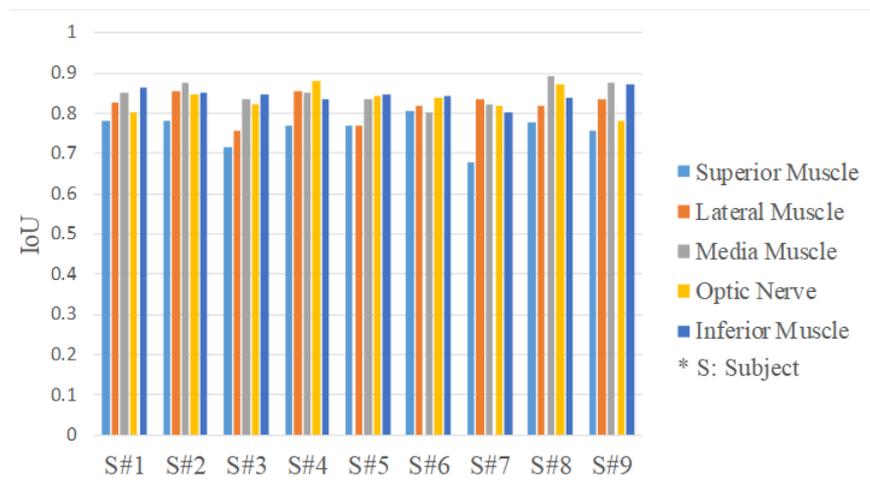

Figure 4. The IoUs of the five target regions in all the 9 test subjects

Table 2 shows the results of Pearson correlation analysis between the volumes calculated from the prediction results and the corresponding ground truth in all the 9 test subjects. All of the R values are greater than 0.98 (P<0.001), which indicates an excellent agreement.

Table 2 Quantitative comparisons between the volumes calculated from the prediction results and the corresponding ground truth. MAE, mean absolute error; RMSE, root mean square error; RE, relative error.

| Target ＼ Metric | Volume in ground truth (cm³) | MAE (cm³) | RMSE (cm³) | R* | RE(%) (min,max) |
|---|---|---|---|---|---|
| Superior rectus Muscle | 0.9977±1.0234 | 0.0845 | 0.1451 | 0.986 | (-14.88,28.27) |
| Lateral rectus Muscle | 0.8419±1.1100 | 0.0802 | 0.1554 | 0.990 | (-11.57, 11.46) |
| Medial rectus Muscle | 0.9543±0.9583 | 0.0677 | 0.1463 | 0.993 | (-6.02, 8.55) |
| Inferior rectus Muscle | 1.1061±1.3280 | 0.0772 | 0.1779 | 0.994 | (-10.23, 11.52) |
| Optic Nerve | 0.4958±0.6933 | 0.0310 | 0.1526 | 0.997 | (-4.90, 10.02) |

*P<0.0001 for R value in all targets (superior rectus muscle, lateral rectus muscle, medial rectus muscle, inferior rectus muscle and optic nerve).

To further demonstrate the performance of our proposed model, we visualize the relative errors for the 5 target regions of all test subjects, as shown in Figure 5.

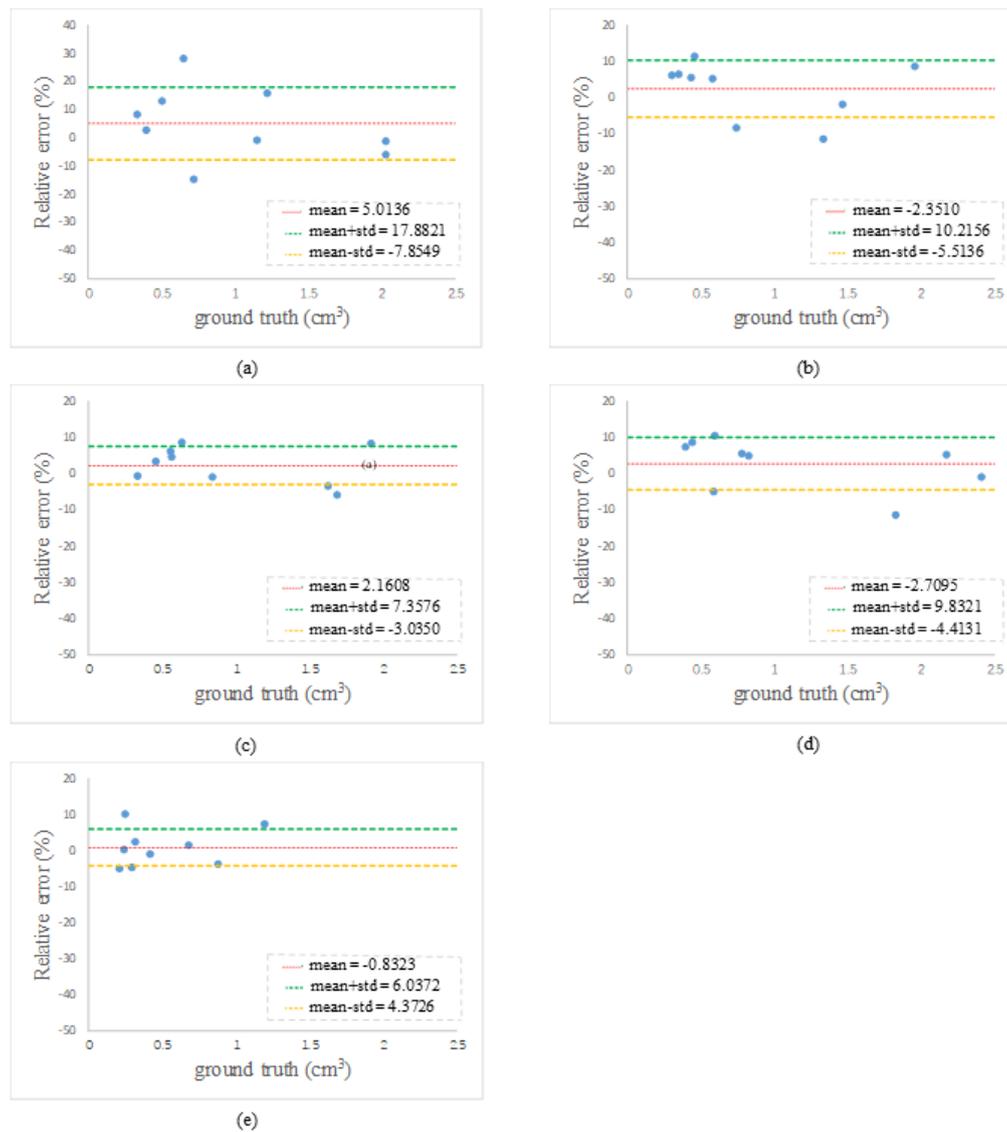

Figure 5. Relative errors between volumes of the segmentation results and ground truth for (a) superior rectus muscle, (b) lateral rectus muscle, (c) medial rectus muscle, (d) inferior rectus muscle and (e) optic nerve.

## 4. Discussion

In this study, we proposed a novel semantic segmentation model to extract the total EOMs and ON from orbital CT images. The overall mIoU, SN, and SP were 0.8207, 0.9129 and 0.9996, respectively. There was an excellent agreement between the volumes calculated from our prediction results and the ground truth (all R>0.98 and P<0.001).

## 4.A. Semantic Segmentation

According to the quantitative comparison in Table 1, SV-net obtained the lowest IoU of 0.7415 in the superior rectus muscle, which presents a relatively inferior performance.

To further investigate this problem, the slice-by-slice IoUs of superior rectus muscle in Subject 7 are shown in Figure 6. At the beginning, the IoU rapidly increases with the slice index. However, in the first few slices the model fails to segment the superior rectus muscle and the IoUs are 0.00.

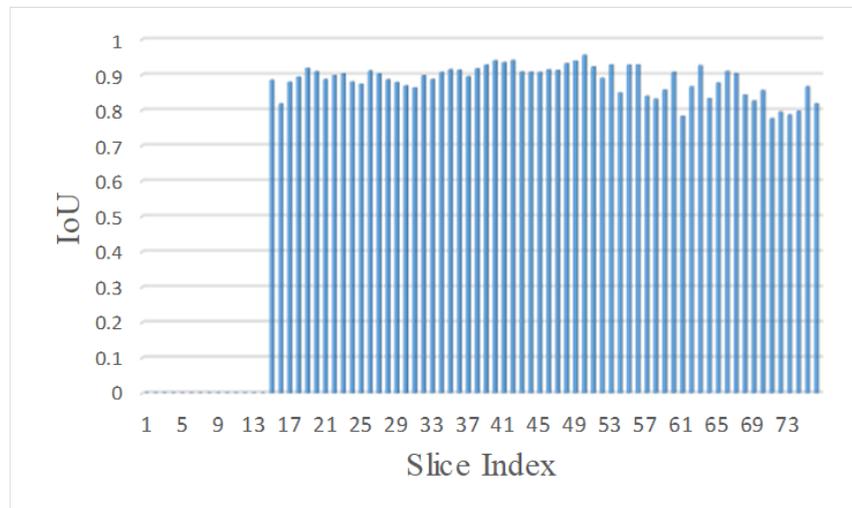

Figure 6. Slice-by-slice IoUs of superior recuts muscle in Subject 7.

Figure 7 visualizes the segmentation results of the first few slices. In the images from 3 to 5, the segmentation results of medial rectus muscle are almost identical with the ground truth; however, the predicted superior rectus muscles are false positive (FP). We hypothesize that the beginning superior rectus muscles are hidden by levator palpebrae superioris so that they are difficult to be separated and contours become unclear[19]. However, because the medial rectus muscle is not hidden by any tissues, it is accurately segmented. The reduced contrast in superior rectus muscle negatively impacts not only the annotations but also the segmentation performance of our proposed model. As a result, there was a relatively inferior segmentation performance for superior rectus muscle. Noteworthy, the above problems caused by hidden tissues commonly occur in the first few slices. A major advantage of our SV-net model is the utilization of the contextual information from neighboring slices, but the ambiguities due to the hidden tissues influence the prediction of inferior rectus muscle within neighboring slices. For example, in the segmentation of the inferior rectus muscle, our SV-net model generated the FP prediction in Slices 7 and 8, and the false negative (FN) prediction with a large portion in Slice 9, as shown in Figure 7.

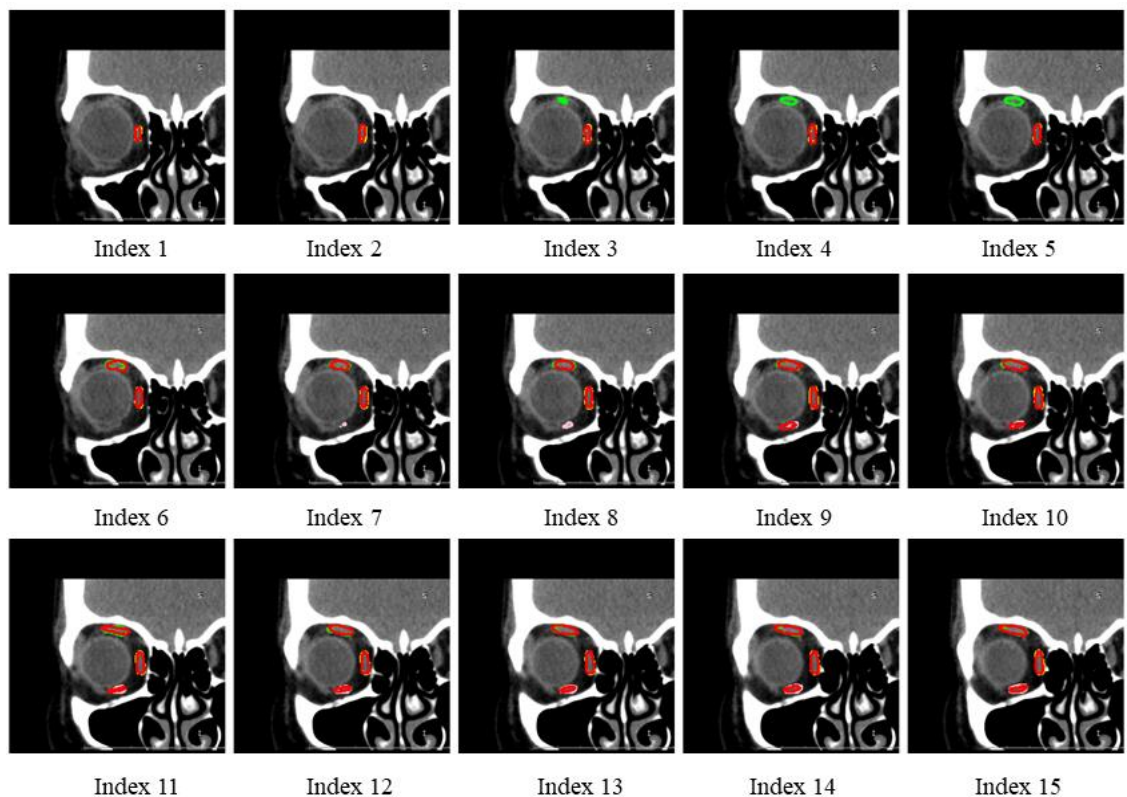

Figure 7. Segmentation results of the first 15 slices in Subject 7. Colors in the annotations reflect the contours of segmentation results by the SV-net: red, ground truth; green, superior rectus muscle; blue, lateral rectus muscle; yellow, medial rectus muscle; pink, inferior rectus muscle; white, optic nerve.

To interpret this phenomenon from a global perspective, Figure 8 demonstrates 3D difference maps of the segmentation results generated by SV-net for five subjects in coronal, sagittal and axial views. In this figure, most of the true positive (TP) voxels (in blue) are wrapped by a small number of FP voxels (in red) and FN voxels (in green), indicating that the extracted contours are close to the ground truth. From the sagittal view, it can be observed that there are more FN voxels (in green) and the FP voxels (in red) in the area far from the fundus oculi (covered by a rectangle) than them in the area near the fundus oculi. It suggests that in the first few slices, the segmentation results were relatively inferior, which can be confirmed in the coronal view as well.

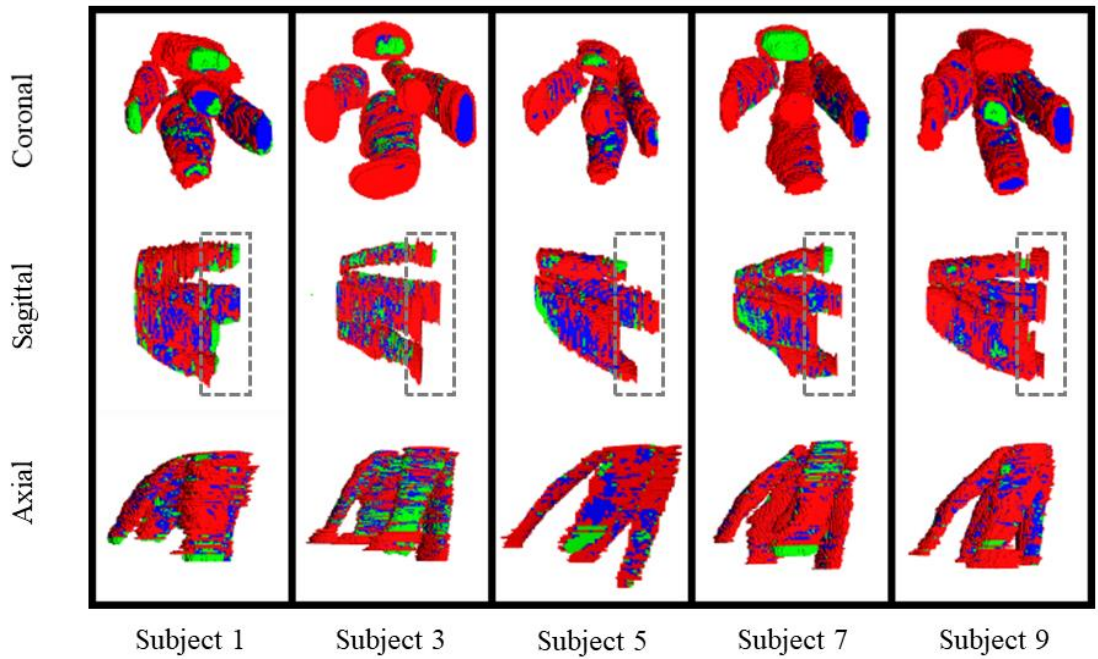

Figure 8. Visualization of 3D difference maps in coronal, sagittal and axial views for 5 subjects. The blue, red and green regions represent the TP voxels, FP voxels and FN voxels, respectively. TP, true positive; FP, false positive; FN, false negative.

## 4.B. Measurement of Volumes

From Table 2, it can be observed that there was an excellent agreement between the volumes calculated from our prediction results and the ground truth (all R>0.98 and P<0.001). Corresponding to the interpretation in the above section, the superior rectus muscle had maximum errors, and its MAE was 0.0845 cm$^3$. Nevertheless, for superior rectus muscle, the Pearson correlation coefficient of the volumes between our prediction and the ground truth was 0.986 (P<0.001), and the relative error was as small as 5.0136%.

## 4.C. Clinical Significance

There was an excellent agreement between our semantic segmentation results and the ground truth. Orbital CT can clearly show the anatomical structure of the EOM and ON and could be used for the quantitative analysis. The automatic extraction and quantification make it easier to assess disease activity and assist in clinical decision making for TAO patients. Additionally, the automatic segmentation of orbital structures could save the operating time and improve the inter- and intra- reproducibility.

Recently, a hybrid imaging method, single photon emission computed tomography (SPECT)/CT was introduced to provide both structural and functional information, which can be used as a referential diagnostic procedure in TAO[20]. Further study will focus on the integration among the automatic calculation of orbital metrics from CT and the functional parameters from SPECT, which will shed light on the progression of particular eye diseases, such as TAO.

## 4.D. Limitation

First, the sample size of our retrospective study is relatively small. Second, as a retrospective study, there was no prior protocol for image acquisition, and the slice thickness is not fixed. Our research is still preliminary, and prospective studies with good quality control in a larger population are needed.

## 5. Conclusions

The qualitative and quantitative evaluations demonstrate excellent performance of our method in automatically extracting the total EOM and ON and measuring their volumes in orbital CT images. There is a great promise for clinical application to assess these anatomical structures for the diagnosis and prognosis of TAO.

### Conflict of interest

The authors declare no conflicts of interest.

### Acknowledgment

This work was supported by National Natural Science Foundation of China (Project Number: 81901784), Henan Science and Technology Development Plan 2020 (Project Number: 202102210384), and Zhengzhou University of Light Industry (Project Number: 2019ZCKJ228). This research was also in part supported by a new faculty startup grant from Michigan Technological University Institute of Computing and Cybersystems.

### References:

1. Forbes G, Gorman CA, Brennan MD, Gehring DG, Ilstrup DM, Earnest Ft. Ophthalmopathy of Graves' disease: computerized volume measurements of the orbital fat and muscle. *AJNR American journal of neuroradiology*. 1986;7(4):651-656.

2. Bahn RS. MECHANISMS OF DISEASE Graves' Ophthalmopathy. *N Engl J Med*. 2010;362(8):726-738.

3. Wiersinga WM, Regensburg NI, Mourits MP. Differential involvement of orbital fat and extraocular muscles in graves' ophthalmopathy. *European thyroid journal*. 2013;2(1):14-21.

4. Jiang CZ, Zhao M, Chen J, et al. Tc-99m DTPA orbital SPECT/CT guided Selection for riamcinolone Acetonide Injection in Patients with Thyroid Associated Ophthalmopathy. *J Nucl Med*. 2019;60:1.

5. Zeng L, Xie XQ, Li CH, Shi HS, Wang F. Clinical study of the radiotherapy with EDGE accelerator in the treatment of the moderate and severe thyroid associated ophthalmopathy. *Eur Rev Med Pharmacol Sci*. 2019;23(8):3471-3477.

6. Comerci M, Elefante A, Strianese D, et al. Semiautomatic Regional Segmentation to Measure Orbital Fat Volumes in Thyroid-Associated Ophthalmopathy: A Validation Study. *The Neuroradiology Journal* 2013;26(4):373-379.

7. Lv B, Wu TN, Lu K, Xie Y. Automatic Segmentation of Extraocular Muscle Using Level Sets Methods with Shape Prior. *International Federation for Medical and Biological Engineering (WC 2012)*. 2013;39(1):904-907.

8. Xing Q, Li Y, Wiggins B, Demer JL, Wei Q. Automatic Segmentation of Extraocular Muscles Using Superpixel and Normalized Cuts. *Advances in Visual Computing, Pt I (Isvc 2015)*. 2015;9474:501-510.

9. Ronneberger O, Fischer P, Brox T. U-Net: Convolutional Networks for Biomedical Image Segmentation. *Medical Image Computing and Computer-Assisted Intervention, Pt Iii*. 2015;9351:234-241.

10. Milletari F, Navab N, Ahmadi S-A. V-Net: Fully Convolutional Neural Networks for Volumetric Medical Image Segmentation. 2016 Fourth International Conference on 3D Vision (3DV); 2016.

11. Russell BC, Torralba A, Murphy KP, Freeman WT. LabelMe: A database and web-based tool for image annotation. *Int J Comput Vis*. 2008;77(1-3):157-173.

12. Li J, Si YJ, Xu T, Jiang SB. Deep Convolutional Neural Network Based ECG Classification System Using Information Fusion and One-Hot Encoding Techniques. Math Probl Eng. 2018;2018:10.

13. Wang T, Lei Y, Tang H, et al. A learning-based automatic segmentation and quantification method on left ventricle in gated myocardial perfusion SPECT imaging: A feasibility study. *Journal of nuclear cardiology. Cardiology.* 2019.

14. Zhao C , Keyak JH, Tang JS, et al. A Deep Learning-Based Method for Automatic Segmentation of Proximal Femur from Quantitative Computed Tomography Images. arXiv preprint arXiv:2006.05513.

15. Sergey Ioffe, Christian Szegedy. Batch Normalization: Accelerating Deep Network Training by Reducing Internal Covariate Shift. ICMI; 2015.

16. Asundi A, Wensen Z. Fast phase-unwrapping algorithm based on a gray-scale mask and flood fill. Applied optics. 1998;37(23):5416-5420.

17. Chang C-I, Wang Y, Chen S-Y. Anomaly Detection Using Causal Sliding Windows. Ieee Journal of Selected Topics in Applied Earth Observations and Remote Sensing. 2015;8(7):3260-3270.

18. Kingma DP, Ba JL. ADAM: A METHOD FOR STOCHASTIC OPTIMIZATION. Computer Science. 2014.

19. Das T, Roos JCP, Patterson AJ, Graves MJ, Murthy R. T2-relaxation mapping and fat fraction assessment to objectively quantify clinical activity in thyroid eye disease: an initial feasibility study. Eye. 2019;33(2):235-243.

20. Szumowski PM, Abdelrazek S, Zukowski, et al. Efficacy of Tc-99m-DTPA

SPECT/CT in Diagnosing Orbitopathy in Graves' Disease. Eur J Nucl Med Mol Imaging. 2018;45:S420-S421.

Table 1 Mean, SD, minimum and maximum of IoU, SP and SN in the nine test subjects predicted by the 10 models from 10-fold cross-validation. SD, standard deviation; IoU, intersection over union; SP, specificity; SN, sentivity.

| Mean±SD (Min,Max) | | Superior Rectus Muscle | Lateral Rectus Muscle | Medial Rectus Muscle | Inferior Rectus Muscle | Optic Nerve | Total(Mean) |
|---|---|---|---|---|---|---|---|
| **SU-net** | IoU | 0.7307±9.73E-05 (0.7142,0.7488) | 0.7968±5.83E-05 (0.7848,0.8083) | 0.8180±2.83E-05 (0.8103,0.8254) | 0.8229±1.99E-05 (0.8148,0.8291) | 0.8090±1.77E-04 (0.7737,0.8201) | 0.7955±2.56E-05 (0.7846,0.8025) |
| | SN | 0.8605±6.67E-04 (0.8064,0.9089) | 0.8877±3.41E-04 (0.8533,0.9153) | 0.9134±2.05E-04 (0.8968,0.9357) | 0.9064±1.88E-04 (0.8885,0.9351) | 0.8866±1.95E-04 (0.8579,0.9106) | 0.8909±1.82E-04 (0.8615,0.9154) |
| | SP | 0.9993±1.81E-08 (0.9990,0.9995) | 0.9996±6.21E-09 (0.9995,0.9998) | 0.9996±5.41E-09 (0.9995,0.9997) | 0.9996±7.97E-09 (0.9994,0.9997) | 0.9998±2.10E-09 (0.9997,0.9999) | 0.9996±4.29E-09 (0.9994,0.9997) |
| **SV-net** | IoU | 0.7415±7.84E-05 (0.7278,0.7599) | 0.8178±3.56E-05 (0.8052,0.8267) | 0.8387±3.62E-05 (0.8267,0.8481) | 0.8383±3.39E-05 (0.8230,0.8436) | 0.8233±9.96E-05 (0.8008,0.8351) | 0.8119±3.22E-05 (0.7997,0.8207) |
| | SN | 0.8937±2.14E-04 (0.8634,0.9159) | 0.9016±2.28E-04 (0.8719,0.9243) | 0.9216±1.02E-04 (0.9041,0.9403) | 0.9221±1.17E-04 (0.9035,0.9374) | 0.9132±5.25E-05 (0.9007,0.9237) | 0.9104±3.02E-05 (0.8981,0.9170) |
| | SP | 0.9992±1.66E-08 (0.9990,0.9994) | 0.9997±1.91E-09 (0.9996,0.9998) | 0.9996±1.30E-09 (0.9996,0.9997) | 0.9996±5.26E-09 (0.9994,0.9997) | 0.9998±9.18E-10 (0.9997,0.9998) | 0.9996±1.40E-09 (0.9995,0.9996) |